\documentclass[journal]{IEEEtran}
\usepackage[utf8]{inputenc}
\author{hak369 }
\date{November 2020}
\usepackage{verbatim}
\usepackage{optidef}
\usepackage{amssymb}
\usepackage[tableposition=top]{caption}
\usepackage{diagbox}

\let\llncssubparagraph\subparagraph
\let\subparagraph\paragraph
\usepackage[compact]{titlesec}
\let\subparagraph\llncssubparagraph

\usepackage{cite}
\ifCLASSINFOpdf
  \usepackage[pdftex]{graphicx}
  
  \DeclareGraphicsExtensions{.pdf,.jpeg,.png}
\else
 
  \usepackage[dvips]{graphicx}

 \DeclareGraphicsExtensions{.eps}
\fi
\usepackage{amsmath}
\usepackage{amssymb}
\usepackage{multirow}
\usepackage{array}
\usepackage{enumitem}
\newcolumntype{P}[1]{>{\centering\arraybackslash}p{#1}}
\newcolumntype{M}[1]{>{\centering\arraybackslash}m{#1}}
\titlespacing*{\section}
{0pt}{0pt}{0pt}
\titlespacing*{\subsection}
{0pt}{0pt}{0pt}
\titlespacing*{\subsubsection}
{0pt}{0pt}{0pt}
\usepackage{xcolor}
\begin{document}

\title{Electricity Tariff Design via Lens of Energy Justice}

\author{Hafiz~Anwar~Ullah~Khan,~\IEEEmembership{Student Member,~IEEE}, Burcin Unel, and Yury Dvorkin,~\IEEEmembership{Member,~IEEE}
  \vspace{-30pt}    
}

\maketitle

\begin{abstract}

Distributed Energy Resources (DERs) can significantly affect the net social benefit in power systems, raising concerns pertaining to distributive justice, equity, and fairness. Electricity tariff and DERs share a symbiotic relationship whereby the design of the former directly impacts the economic efficiency and equity in the system. Current tariff design approaches suffer from opaque efficiency-equity trade-offs and are also agnostic of the externalities that affect both economic efficiency and equity. Therefore, this paper develops a justice-cognizant tariff design framework that improves the economic efficiency of tariff without sacrificing its distributional equity, and encompasses economic welfare, social costs of environmental and  public health impacts, and socio-economic and demographic characteristics of electricity consumers. The proposed framework is based on a Single Leader Single Follower (SLSF) game incorporating a multi-objective optimization problem, and is evaluated on four different tariff structures. The SLSF game is reformulated as a Multi-Objective Problem with Equilibrium Constraints (MOPEC) and is solved by integrating the objective sum method for multi-objective optimization and Scholtes’s relaxation technique for equilibrium constraints. We compare the economic efficiency and equity of the proposed framework using the 11-zone New York ISO and 7-bus Manhattan power networks. The results demonstrate that spatially- and temporally-granular tariffs ensure equity and economic efficiency at a lower energy burden to consumers. 

\end{abstract}

\IEEEpeerreviewmaketitle
\section*{Nomenclature}
\vspace{-5pt}
\subsection{Sets and Indices}
\begin{IEEEdescription}[\IEEEusemathlabelsep\IEEEsetlabelwidth{$\overline{\alpha}_{t,b},\underline{\alpha}_{t,b}$}]
\item[$b_0$] {Root node of the distribution system}
\item[$B_{m/n} (b)$] {Set of ancestor/children buses of bus $b$ }
\item[$b^\textnormal{T}_c$]{Bus in transmission network connected to $b_0$ }
\item[\textbf{$b \in B$}] {Set of buses in the distribution network}
\item[$i \in I$]{Set of generators in distribution networks}
\item[$r \in R$]{Set of representative operating days}
\item[$T_1/T_2, (T)$] {Set of time intervals in peak/off-peak demand time, ($T_1 \cup T_2$), indexed by $t$ }
\item[$y \in \Theta$] {Set of air pollutants under consideration}
\end{IEEEdescription}
\vspace{-5pt}
\subsection{Parameters}
\begin{IEEEdescription}[\IEEEusemathlabelsep\IEEEsetlabelwidth{$\overline{\alpha}_{t,b},\underline{\alpha}_{t,b}$}]
\item[$C_{i}$] {Operational cost of generator $i$ }  
\item[$C^{\textnormal{cap}}$]{Capital cost of utility (prorated on daily basis)}
\item[$D^\textnormal{{p}}_{b,t,r}$] {Total inflexible active power demand at bus $b$}
\item[$d^\textnormal{{q}}_{b,t,r}$] {Inflexible reactive power demand at bus $b$}
\item[$G^{{\textnormal{max/min}}}_{i}$] {Max/min power output of generator $i$}
\item[$H_{b}$]{Average size of household at bus $b$}
\item[$P_{b}$] {Total population at bus $b$}
\item[$Q^{{\textnormal{max/min}}}_{i}$] {Max/min reactive power output of generator $i$}
\item[$R_{y,i}$] {Emission factor of generator $i$ for pollutant $y$} 
\item[$S_{(b,b_1)}$] {Maximum apparent power flow in distribution feeder between $b$ and $b_1 \in B$}
\item[$U^{{\textnormal{max/min}}}_{b}$] {Max/min square of voltage magnitude at bus $b$}
\item[$X_{(b,b_1)}$] {Resistance of line connecting $b$ and $b_1 \in B$}
\item[$x_{(b,b_1)}$] {Reactance of line connecting $b$ and $b_1 \in B$}
\item[$\alpha_{b}$]{Elasticity of demand at bus ${b}$}
\item[$\gamma$] {Carbon tax imposed by the regulator}
\item[$\gamma^\textnormal{sc}$] {Social cost of CO\textsubscript{2} emissions}
\item[$\kappa/\kappa'$]{Economic burden percentage set by the regulator/household}
\item[$\lambda_{y,b}$]{External cost of pollutant $y \in \Theta$ at bus $b$ }
\item[$\pi_{b}^{\textnormal{min/max}}$] {Min/Max tariff value for bus $b$}
\item[$\pi^{\textnormal{avg}}$]{Average value of tariff set by the regulator}
\item[$\mu^\textnormal{inc}_{b}$]{Average household income at bus ${b}$}
\item[$\nu$]{Parameter relating peak and off-peak tariffs}
\item[$\upsilon$]{Rate of return for power utility}
\end{IEEEdescription}

\subsection{Variables}
\begin{IEEEdescription}[\IEEEusemathlabelsep\IEEEsetlabelwidth
{$\overline{\alpha}_{t,b},\underline{\alpha}_{t,b}$}]
\item[$d^\textnormal{{p}}_{b,t,r}$] { Flexible active power demand at bus $b$}
\item[$d^\textnormal{{p}}_{b,T_{1}/T_2,r}$] {Flexible active power demand at bus $b$ for $T_1/T_2$}
\item[$e^{\textnormal{D}}_y$]{Emission $y \in \Theta$ in distribution network}
\item[$e^{\textnormal{total}}_y$]{Total system-wide value of pollutant $y \in \Theta$}
\item[$e^{\textnormal{D}}_{y,b}$]{Emission $y$ in distribution network at bus $b$}
\item[$f^{\textnormal{EW/H/EN}}$]{Objective function associated with economic welfare/health/environment}
\item[$f^{\textnormal{p/q}}_{(b,b_m),t,r}$] {Active/reactive power flow in line b/w $b$ and $b_m$}
\item[$g_{b,t,r}$] {Power output of generator at bus $b$}
\item[$g_{i,t,r}$]{Power output of generator $i$}
\item[$(g/q)_{b_0,t,r}$] {Active/reactive power output of generator at $b_0$}
\item[$q_{b,t,r}$] {Reactive power output of generator at bus $b$}
\item[$u_{b,t,r}$] {Square of voltage magnitude at bus $b$}
\item[$\lambda^{\textnormal{T}}_{b^{\textnormal{T}}_c,t,r}$] {Wholesale market-clearing price at bus $b^{\textnormal{T}}_c$}
\item[$\pi_{b,t,r}$] {Energy tariff for bus $b$ at time $t$}
\item[$\pi_{b,t,r}^\textnormal{p/op}$] {Energy tariff for bus $b$ during peak/off-peak time intervals $t \in T_1$ / $t \in T_2$}
\end{IEEEdescription}

\section{Introduction}

Integration of distributed energy resources (DERs) into the distribution network offers significant benefits to electricity consumers, utilities, and alternative service providers (aggregators/retailers) \cite{nature_galina}. From the perspective of consumers, DERs reduce costs and emissions and increase the reliability of electricity, whereas from the perspective of utilities and alternative service providers, DERs provide new ways to offer better services to the consumers \cite{Fares}. 
However, in case of poor integration, e.g. `connect and forget' approach \cite{IRENA}, DERs decrease the net social benefit and increase inequity in the system by drastically increasing costs and emissions \cite{utility_mit, nature_Brockway}. Hence, regulators, utilities, consumer advocates, and policymakers all seek DER integration solutions that ensure an equitable distribution of costs and benefits in the power system. Any such solution will require extensive changes in policy and electric power distribution regulation, e.g. in the design of carbon tax, renewable energy credits \cite{RECs}, incentives for installing and operating Renewable Energy Sources (RES), and renewable portfolio standards \cite{RPS}. However, a DER-centric  overhaul of current tariff design practices forms the core of these policy changes \cite{utility_mit}. 

In practice, the design of electricity tariffs is not only a techno-economic problem, but social and political acceptability of the designed tariff is also important for its adoption. A key consideration is the concept of equity/fairness in tariff design, as claimed for instance in the tariff design principles of the Massachusetts Department of Public Utilities \cite{Mass_DPU}, New York Department of Public Service \cite{NY_DPS}, and Nevada Public Utilities Commission \cite{Nevada}. For example, in 2017, 45 states in the U.S. proposed or adopted changes to tariff design practices  to enable such DER integration that increases the net social benefit of the power system \cite{NC_Q4}. This equity-oriented overhaul to tariff design practices shows that regulators in power systems aim to design electricity tariff such that the benefits of electricity among members of society are distributed equitably. However, existing tariff design practices are marred by a misleading dichotomy of economic efficiency versus equity, resulting in economically inefficient and socially inequitable tariffs \cite{tariff_DER_MIT, Burcin_RateDesign}. Therefore, it is critical to formulate a modeling framework for electricity tariff design that not only considers efficiency and ensures techno-economic feasibility but also recognizes and ameliorates  social inequities in the current electric power infrastructure. We underscore that our results, obtained using real-life data from New York, NY, reveal that economic efficiency of electricity tariffs can be achieved without sacrificing equity.

\subsection{Literature Review}
Design of electricity tariff to enable proper integration of DERs into the power system has been widely discussed in literature \cite{tariff_DER_MIT, Burcin_RateDesign}. Recent work on tariff design focuses on power systems with a high RES penetration, predominantly residential rooftop PVs \cite{Bloch}. For instance, Ansarin \textit{et al.} evaluate the fairness and efficiency of five different tariff structures including conventional tariff, flat-rate tariff, time-of-use (TOU) tariff, real-time pricing, and demand charge tariff \cite{Ansarin}. The results show that flat volumetric tariffs perform poorly in terms of economic efficiency and equity, allowing wealth transfer and cross-subsidization between non-DER-owners to DER owners. However, time-based tariffs are more economically  efficient and equitable. Burger \textit{et al.} analyze the changes in electricity bill among various socio-economic groups in Chicago for different tariff structures. They conclude that regulators can improve the economic efficiency of tariffs without sacrificing distributional equity if proper changes are introduced to fixed charges for recovering the system residual costs \cite{Burger}. Similarly, \cite{Burcin, Burcin_emissions} simulate the effect of various pricing mechanisms on household adoption of DERs in the service territory of Commonwealth Edison in Chicago. Six different pricing mechanisms pertaining to flat volumetric tariffs and system-cost reflective tariffs are simulated. Again, the results demonstrate that adoption of DERs with time-based system-cost reflective tariffs  leads to greater reduction in peak demands as compared to flat volumetric tariffs. Additionally, tariff optimization problem has also been studied for designing TOU tariff \cite{Hung}, demand response management \cite{Kovacs}, congestion management \cite{Shen}, and RES-based nano-grid design for dynamic tariff \cite{Parizy}. Other tariff structures that extend and complement the traditional TOU structure, including prediction-of-use tariff \cite{Robu} and time-and-level-of-use tariff \cite{Besancon},  are also discussed in literature. Notably models and analyses in \cite{Burcin, Burcin_emissions,Bloch, Kovacs, Shen, Parizy, Oprea, Hung, Robu, Besancon} do not consider equity/fairness in  tariff design. 

While the current literature delves in depth into the techno-economic design of electricity tariffs, an evident research gap exists in quantifying the trade-off between economic efficiency and equitability of electricity tariff. To the best of authors' knowledge, while the existing literature provides qualitative foundation for equity in tariff design \cite{tariff_DER_MIT}, and uses limited data-based approaches to quantify fairness \cite{Ansarin, Burger}, a comprehensive framework for internalizing equity/fairness considerations in tariff design is missing. In this paper, we aim to bridge this gap by proposing a justice-cognizant tariff design framework that incorporates social and techno-economic considerations of power systems, while encapsulating the economic, public health, and environmental impacts of energy production and consumption.

\begin{figure}[!t]
\centering
\includegraphics[width = 3.5in, height=2.8in]{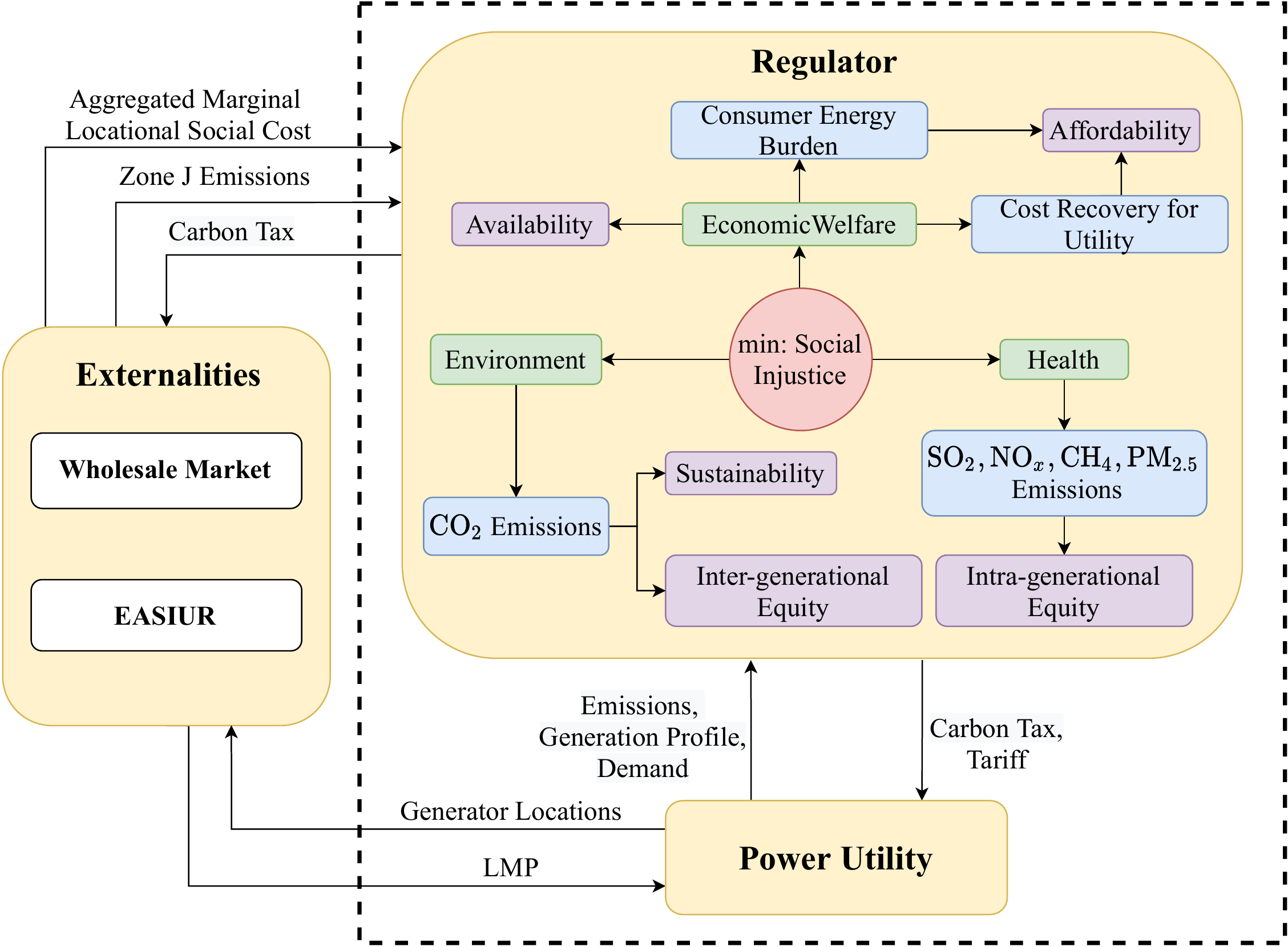}
\caption{\small{A schematic diagram showing the SLSF game between the regulator and the power utility (shown in dashed box). Red color displays the multi-objective function of the regulator, with its individual components shown in green. Blue color shows the constraints, while purple depicts the justice considerations. }}
\vspace{-20pt}
\label{schematic}
\end{figure}

\subsection{Problem Structure and Description}
Currently, tariffs are determined by the public utility commissions (also known as regulators) in the power system. Regulators accept tariff proposals from power utilities and determine tariffs based on techno-economic considerations of the system. These tariffs are then adopted by  both the consumers and the power utilities \cite{utility_mit}. The structure of this problem coincides with the Stackelberg games, where the regulator acts as the leader, and the power utility along with the consumers  are followers. Hence, the justice-cognizant tariff-design framework can be formulated as a Single Leader Single Follower (SLSF) game between  the regulator and  power utility, as shown in Fig~\ref{schematic}. Since the utility also procures power from the wholesale market, we externalize this component of the problem using Locational Marginal Prices (LMPs) at the interconnection buses between the transmission and distribution (T\&D) networks. The market-clearing decisions of the wholesale market - LMP and emissions - are treated as parameters in the designed SLSF game.

To develop a justice-cognizant tariff-design framework based on a SLSF game, we adapt the energy justice framework presented by Sovacool \textit{et al.} \cite{Sovacool_2, Sovacool_3}. Based on \cite{Sovacool_2, Sovacool_3}, we define seven considerations for the proposed justice-cognizant tariff-design framework. These considerations include availability and affordability of energy, environmental sustainability, health impacts, inter- and intra-generational equity, and economic welfare. Fig. \ref{schematic} shows these principles in purple along with the objectives/constraints that are used to represent these principles in the regulator's problem. For example, from the viewpoint of the regulator, we model availability and affordability of energy using power balance and energy burden (EB)  constraints \cite{Sovacool_4}, respectively. Environmental sustainability of the energy produced in response to the demand (created as a function of designed tariff) is evaluated using $\textnormal{CO}_2$ emissions. In addition, we consider the public health impacts of such pollutants as $\textnormal{SO}_2$, $\textnormal{NO}_x$, $\textnormal{CH}_4$, and $\textnormal{PM}_{2.5}$. Unlike $\textnormal{CO}_2$, which is a pollutant with global damages, these pollutants have local damages. Their impact on population health varies with the location at which these pollutants are emitted. The locational changes in external costs are a function of  local pollutant concentrations. Hajat \textit{et al.} in \cite{air_pollution_rev} conclude that socio-economically marginalized communities in North America experience higher  pollutant concentrations. Therefore, a uniform social marginal cost of these pollutants cannot be used for the entire system. This motivates the use of variable resolution grids in air quality models that employ high spatial resolution (e.g. few kilometers) in high-population density areas \cite{EASIUR, INMAP}. Accordingly, we use the Estimating Air pollution Social Impact Using Regression (EASIUR) model \cite{EASIUR} to estimate the locational marginal external costs (or public health costs) of local pollutants in the system. This model reduces the computational complexity of the conventional chemical transport models and uses regression to calculate the external costs of emissions based on the geographical location of their source \cite{EASIUR}. EASIUR is external to the designed SLSF game, as shown in Fig.\ref{schematic}, such that the locations of generators from the wholesale market and the power utility serve as input to the EASIUR model. The output of EASIUR is the locational marginal external cost of aggregated generators on each bus of the power utility, which is then communicated to the regulator.  Similarly, inter-generational equity is defined in terms of minimizing the impact of global warming potential of $\textnormal{CO}_2$ here-and-now to reduce  the impact of current energy production passed on to the future generations. Health impacts of local pollutants, as described above, are also used as a proxy for modeling intra-generational equity such that the health costs of energy production are equitably distributed in the system. Finally, economic welfare is formulated using the utility functions and costs of consumers, and revenue and operational/capital costs of the power utility. We use economic welfare, health, and environmental sustainability in the objective function of the regulator, whereas availability and affordability along with revenue adequacy considerations are modeled as constraints in the regulator's problem. Hence, we develop a SLSF Stackelberg game incorporating a multi-objective problem for the leader (regulator) and a  conventional profit-seeking problem for the follower (power utility).

\subsection{Solution Techniques for Stackelberg Games}
One of the recurring strategies to solve Stackelberg games leverages KKT conditions or duality theory to convert the bi-level problem into a single-level equivalent. To avoid nonlinearities introduced by the application of duality theory and dealing with their mixed-integer linear reformulations needed for employing off-the-shelf solvers, we use KKT conditions in this paper. The KKT reformulation introduces complementarity constraints, leading to a Mathematical Program with Equilibrium Constraints (MPEC). However, our upper-level (UL) problem in the formulated SLSF game, shown in Fig.~\ref{schematic} as the regulator, is a multi-objective problem, due to which the resulting single-level equivalent is a Multi-Objective Problem with Equilibrium Constraints (MOPEC), instead of an MPEC. 

Multiple solution methodologies to solve multi-objective optimization problems are available in literature that vary based on the articulation of preferences by the decision-maker \cite{MOOP}. These solution techniques include the objective sum method, min-max method \cite{min_max}, objective product method, Nash arbitration method, and etc. \cite{MOOP}. Min-max method provides sufficient conditions only for a weak Pareto optimal point, whereas the objective product method and Nash arbitration method introduce nonlinearities in the problem \cite{MOOP}. Therefore, in this paper, we use the objective sum method to treat the multi-objective function because it provides sufficient conditions for optimality without introducing extra complexity to multi-objective problems \cite{MOOP}. However,  this method can only be used when articulation of preferences among individual objectives in the multi-objective function is a priori known to the policymaker. Moreover, due to the existence of complementarity constraints in MOPECs, standard Mangasarian-Fromovitz constraint qualification does not hold. Therefore, convergence assumptions of standard nonlinear programs (NLPs) are invalid for MOPECs. Hence, weaker stationarity concepts, such as C, M, B, and strong stationary points are defined in this case \cite{MOPEC_stat}.  The convergence of the obtained solution to one of these points defines the proximity of this solution to the actual optimizer \cite{relaxation_comp}. Khan \textit{et al.} \cite{SLMF} discuss solution methodologies for treating complementarity constraints in MPECs (and by extension MOPECs), and state that relaxation of complementarity constraints makes the problem amenable to be solved using off-the-shelf NLP solvers. Various relaxation approaches are proposed in literature to enhance the theoretical properties (such as convergence to a stronger stationary point) of the relaxed NLPs \cite{SLMF}. However, numerical results demonstrate that the proposed relaxation techniques do not converge to the theoretically claimed stronger stationary point, rather converge to an inexact one, leading to a weaker stationarity result. The relaxation scheme by Scholtes \cite{Scholtes}, on the contrary, converges to a relatively strong C-stationary point, even if the convergence is obtained to an inexact stationary point \cite{inexact}. This characteristic is thus superior to other proposed relaxation methodologies and used for solving MOPECs in this paper. 

Hence, we develop an integrated solution methodology to treat the multi-objective function in the UL problem using the objective sum method, while simultaneously treating complementarity constraints with Scholtes's relaxation technique in the lower-level (LL) problem. The proposed solution technique guarantees convergence to the strongest achievable stationary point using off-the-shelf NLP solvers.

\subsection{Contributions}
The main contributions of this paper are: 
\begin{enumerate}
    \item  We propose a justice-cognizant tariff-design framework that models the techno-economic and equity considerations in the system.  Equity is modeled using the locational costs of public health from EASIUR, global climate change damages, and  socio-economic and demographic profiles of consumers.
    \item We formulate a SLSF game between the regulator and the power utility to determine justice-aware electricity tariffs. This game is solved using the integrated solution methodology for MOPECs employing the objective sum method and Scholtes's relaxation scheme. 
    \item The case study compares four different tariffs, where values of the EB for consumers that result in optimal justice-cognizant tariff in each case are provided. The results show that economic efficiency of tariff can be achieved without sacrificing equity in the system.
\end{enumerate}

\section{Problem Formulation}

This section describes a mathematical formulation of each player in the proposed SLSF game. 

\subsection{Formulation of the Regulator}
\label{Sec:Form_Reg}
The regulator internalizes social factors into a tariff design optimization by means of leveraging the multi-objective function stated in eq.~\eqref{r1}, which maximizes economic welfare ($f^\textnormal{EW}$), and minimizes the impact of energy production on public health ($f^\textnormal{H}$) and environmental  ($f^\textnormal{EN}$), using the electricity tariff as the decision variable. The economic welfare, defined in eq.~\eqref{r2}, consists of the revenue of the power utility, surplus of the consumers, and capital and operating costs of energy production, including the cost of electricity procurement from the wholesale market. Next, the impact of energy production on population health is defined in eq.~\eqref{r3}, where the total value of each pollutant emitted at each bus ($e^{\textnormal{T/D}}_{y,b^\textnormal{T/D}}$) is multiplied by the locational marginal external cost of that pollutant ($\lambda_{y,b^\textnormal{T/D}}$) obtained from EASIUR \cite{EASIUR}. Finally, the environmental damage due to energy production is defined in eq.~\eqref{r4}, where total system-wide CO\textsubscript{2} emissions ($e^{\textnormal{total}}_{CO_2}$) are weighted by their net external cost ($\gamma^{\textnormal{sc}} - \gamma$). Note that the  net external cost ($\gamma^{\textnormal{sc}} - \gamma$) is used instead of the complete Social Cost of Carbon ($\gamma^{\textnormal{sc}}$) to avoid double-counting the environmental damage associated with carbon emissions. Accordingly, the problem of the regulator, for the set of representative days $R$ over which the optimization is performed, can be formalized as follows:

\vspace{-18pt}
\allowdisplaybreaks
\begin{subequations}
\label{regulator}
\begin{align}
\label{r1}
\min_{\pi_{b,t,r}}\hspace{5pt} &\textnormal{O}= [-f^{\textnormal{EW}},f^{\textnormal{H}},f^{\textnormal{EN}}]\\ 
\label{r2}
\begin{split}
&f^{\textnormal{EW}} :=  \underbrace{\sum_{b\in B, t \in T, r \in R}{\pi_{b,t,r}(D^\textnormal{p}_{b,t,r} + d^\textnormal{p}_{b,t,r})}}_\text{Revenue of power utility}\\&\hspace{-10PT} +\Bigg[\hspace{-2pt}\sum_{b \in B, t_1,t_2 \in T_1,T_2, r \in R}\hspace{-30pt} \Bigg.\hspace{-4PT}(d^\textnormal{p}_{b,t_1,r} +D^\textnormal{p}_{b,t_1,r})^{\alpha_{b}} (d^\textnormal{p}_{b,t_2,r}+D^\textnormal{p}_{b,t_2,r})^{1-\alpha_{b}}\\& \underbrace{- \pi_{b,t_1,r}^{\textnormal{p}}(d^\textnormal{p}_{b,t_1,r}+D^\textnormal{p}_{b,t_1,r}) - \pi_{b,t_2,r}^{\textnormal{op}}(d^\textnormal{p}_{b,t_2,r}+D^\textnormal{p}_{b,t_2,r})}_\text{Utility function and cost to consumers}\Bigg. \Bigg]\\&- \underbrace{\sum_{i \in I,t \in T, r \in R}\hspace{-10pt}{C_{i}g_{i,t,r}} - \sum_{t \in T, r \in R} \lambda^\textnormal{T}_{b^\textnormal{T}_c,t,r} g_{b_{0},t,r}}_\text{Cost for generation/procurement of electricity} - {C^\textnormal{cap}}
\end{split}\\
\label{r3}
&f^{\textnormal{H}} := \hspace{-30pt}\underbrace{\sum_{y \in \Theta/CO_2, b \in B}\hspace{-10pt}{e^{\textnormal{D}}_{y,b}\lambda_{y,b}}}_\text{For pollutants emitted in distribution system} \hspace{-5pt} +\hspace{-5pt} \underbrace{\sum_{y \in \Theta/CO_2}{\hat{e}^{\textnormal{T}}_{y,b^\textnormal{T}_c}\lambda_{y,b^\textnormal{T}_c}}}_\text{For pollutants emitted at $b^\textnormal{T}_c$}\\
\label{r4}
&f^{\textnormal{EN}} :=  (\gamma^{\textnormal{sc}} - \gamma) e^{\textnormal{total}}_{CO_2}\\
\text{s.t.}\hspace{15pt}
\label{r6}
& \sum_{t \in T, r \in R}\underbrace{{\left(\frac{d^\textnormal{p}_{b,t,r} + D^\textnormal{p}_{b,t,r} }{P_{b}}\times H_{b}\right)}}_\text{Power demand per household} \pi_{b,t,r} \leq \kappa\mu^\textnormal{inc}_{b}; \hspace{10 pt} \forall b \in B\\
\label{r8}
\begin{split}
& \hspace{-10pt}\sum_{b\in B, t \in T, r \in R}{\pi_{b,t,r}(D^\textnormal{p}_{b,t,r} + d^\textnormal{p}_{b,t,r})} = (1 + \upsilon){C^\textnormal{cap}}  \\& \hspace{20pt}+ \left[\sum_{i \in I,t \in T, r \in R}\hspace{-10pt}{C_{i}g_{i,t,r}} + \sum_{t \in T, r \in R} \hspace{-1 pt}\lambda^\textnormal{T}_{b^{\textnormal{T}}_c,t,r} g_{b_{0},t,r}\right]
\end{split}\\
\label{r9}
& \pi_{b,t}^{\textnormal{min}} \leq \pi_{b,t,r} \leq \pi_{b,t}^{\textnormal{max}}; \hspace{0 pt} \forall b \in B, t \in T, r \in R\\
\label{r10}
& \pi_{b,t,r}^\textnormal{p} \geq \nu \times \pi_{b,t,r}^\textnormal{op}; \hspace{10 pt} \forall b \in B, t \in T, r \in R\\
\label{r12}
&e_{CO_2}^{\textnormal{total}} = \hat{e}_{CO_2,b^\textnormal{T}_c}^\textnormal{T} +e_{CO_2}^D\\
\label{r11}
\begin{split}
&\sum_{t \in T_1, b \in B, r \in R}{\frac{\pi_{b,t,r}^\textnormal{p}}{\mathbf{card}(T_1)\mathbf{card}(b)}} \\&+ \sum_{t \in T_2, b \in B,r \in R}\hspace{-3pt}{\frac{\pi_{b,t,r}^{\textnormal{op}}}{\mathbf{card}(T_2)\mathbf{card}(b)}} \hspace{10pt}\leq 2 \times \pi^{\textnormal{avg}} 
\end{split}\\
\label{tariff_def}
&\pi_{b,t,r} =
\begin{cases}
\pi_{b,t,r}^\textnormal{p}, &\forall t \in T_1, b \in B, r \in R  \\ \pi_{b,t,r}^{\textnormal{op}}, &\forall t \in T_2, b \in B, r \in R
\end{cases}
\end{align}
\end{subequations}

Eq.~\eqref{r6} is the EB constraint, which constrains the daily energy expense for the median household at each bus to be less than a pre-defined percentage ($\kappa$) of the daily mean household income at that bus. The annual household income \cite{income_data} is prorated on a daily basis using a 5\% discount rate. Eq.~\eqref{r8} ensures revenue adequacy for the power utility, i.e. the ability of the utility to recover its capital cost along with a pre-negotiated rate of return ($\upsilon$), and its operating cost from the revenue streams. Eq.~\eqref{r9}  imposes the minimum and maximum tariff limits and  eq.~\eqref{r10} relates  the peak and off-peak tariffs via parameter $\nu\geq1$. Eq.~\eqref{r11} restricts the average value of the peak and off-peak tariffs in the system to a pre-defined value. Since the peak and off-peak tariffs are averaged over both the set of time intervals and set of buses (using cardinality operator of sets $\mathbf{card (.)}$), a factor of 2 is introduced in the right-hand side to ensure accurate averaging of the tariff across two sets.  The total CO\textsubscript{2} emissions in the system are calculated in eq.~\eqref{r12}, whereas eq.~\eqref{tariff_def} defines the time-of-use tariffs for peak and off-peak time intervals. Note that a  system-wide, time-of-use tariff can be modeled if  $\pi_{b,t,r}^\textnormal{p}$ = $\pi_{b,t,r}^{\textnormal{op}} \forall b \in B$. Similarly, enforcing $\pi_{b,t,r}^\textnormal{p}$ = $\pi_{b,t,r}^{\textnormal{op}} \forall b \in B,  t\in T, r \in R$ will lead to the system-wide, time-invariant tariff. In this case, eq.~\eqref{r10} is not imposed.

\subsection{Formulation of the Power Utility}
The objective function of the power utility in eq. \eqref{6} is to maximize its regulated profit, which is defined as the difference between the tariff-based revenue from selling electricity to consumers (where the tariff is optimized by the regulator as explained in Section~\ref{Sec:Form_Reg}) and the operating cost.  In turn, the operating cost includes the production cost of distribution-level generation resources operated by the utility,  the cost of the interface flow  between the wholesale market and power utility, and carbon tax payments. Electric power distribution is modeled using  the \textit{LinDistFlow} AC power flow formulation \cite{lindist} in (\ref{7}) -   (\ref{121}). The nodal active and reactive power balance at the root node of the distribution network are enforced in \eqref{7} and \eqref{8}, while at all other buses in (\ref{9}) and (\ref{91}).  Similarly, capacity constraints for all generating units are modeled in (\ref{10}) and (\ref{14}). Nodal voltage and line power flow limits are enforced in (\ref{12}), \eqref{12.1},  and (\ref{13}). Constraint \eqref{12.1} models the  power exchange limit between the T\&D systems, whereas \eqref{13} constrains  apparent power  flows in distribution lines \cite{lindist}. Note that  \eqref{13} is a conic constraint, where $K := \{x \in \mathbb{R}^3 \vert x_1^2 \geq x_2^2 +x_3^2\}$ and  $K^\ast$  denote primal and dual second-order cones \cite{boyd}.  Nodal voltage magnitudes are modeled in eq.~\eqref{121}. Eq.~\eqref{j4} calculates the emissions produced by the generators at each bus of the system, whereas eq.~\eqref{em_d} converts these emissions to the system-wide emissions for each pollutant.

\vspace{-20pt}
\allowdisplaybreaks
\begin{subequations}
\label{distribution}
\begin{align}
\label{6}
\begin{split}
\max_{\Xi^\textnormal{U}} &\sum_{b\in B, t \in T, r \in R}{\pi_{b,t,r}(d^\textnormal{p}_{b,t,r} + D^\textnormal{p}_{b,t,r})}\\& -\hspace{-10pt} \sum_{i \in I,t \in T, r \in R}\hspace{-15pt}{C_{i}g_{i,t,r}} - \hspace{-10pt}\sum_{t \in T, r \in R} \hspace{-1 pt}\lambda^\textnormal{T}_{b^\textnormal{T}_c,t,r} g_{b_{0},t,r} - \gamma e^{\textnormal{D}}_{CO_2} 
\end{split}\\
\vspace{-8pt}
\text{s.t. \hspace{-3pt}\{}
\vspace{-7pt}
\label{7}
&{g_{b_{0},t,r}} = \sum_{b_n \in B_n(b_{0})}{f^p_{(b_{0},b_n),t,r}}: (\lambda_{b_0,t,r}^\textnormal{D})\\
\label{8}
&{q_{b_{0},t,r}} = \sum_{b_n \in B_n(b_{0})}{f^\textnormal{q}_{(b_{0},b_n),t,r}}: (\lambda_{b_0,t,r}^\textnormal{Dq})\\
\label{9}
\begin{split}
&g_{b,t,r} + \sum_{b_m \in B_m(b)}{f^\textnormal{p}_{(b,b_m),t,r}} = d^\textnormal{p}_{b,t,r} + D^\textnormal{p}_{b,t,r}+\\[-4pt]& \sum_{b_n \in B_n(b)}{f^\textnormal{p}_{(b,b_n),t,r}} :(\lambda_{{b,t,r}}^\textnormal{D}); \hspace{3mm} \forall b \in B
\end{split}\\
\label{91}
\begin{split}
&q_{b,t,r} + \sum_{b_m \in B_m(b)}{f^\textnormal{q}_{(b,b_m),t,r}} = d^\textnormal{q}_{b,t,r}+\\[-4pt]&\hspace{-8pt}\sum_{b_n \in B_n(b)}{f^\textnormal{q}_{(b,b_n),t,r}}: (\lambda_{{b,t,r}}^\textnormal{Dq}); \hspace{5mm} \forall b \in B 
\end{split}\\
\label{10}
  &G_{i}^{{\textnormal{min}}} \leq  g_{i,t,r} \leq G_{i}^{{\textnormal{max}}}:(\underline{\delta}_{i,t,r}, \overline{\delta}_{i,t,r}); \forall i \in I\\
\label{14}
  &Q_{i}^{{\textnormal{min}}} \leq  q_{i,t,r} \leq Q_{i}^{{\textnormal{max}}}:(\underline{\theta}_{i,t,r}, \overline{\theta}_{i,t,r});\forall i \in I\\
 \label{12} 
 &U_{b}^{{\textnormal{min}}} \leq  u_{b,t,r} \leq  U_{b}^{{\textnormal{min}}}:(\underline{\mu}_{b,t,r}, \overline{\mu}_{b,t,r}); \forall b \in B \\
 \label{12.1}
 &\hspace{-15pt}-F_{(b^\textnormal{T}_c,b_0^\textnormal{D})}^{\textnormal{max}}\leq f_{(b^\textnormal{T}_c,b_o^D),t,r} \leq F_{(b^\textnormal{T}_c,b_0^\textnormal{D})}^{\textnormal{max}}\hspace{-4pt}:\hspace{-3pt} (\underline{\tau}_{(b^\textnormal{T}_c,b_0^\textnormal{D}),t,r}, \overline{\tau}_{(b^\textnormal{T}_c,b_0^\textnormal{D}),t,r})\\
 \vspace{-5pt}
\label{13}
\begin{split}
\vspace{-5pt}
&[S_{(b,b_1)}; f^\textnormal{p}_{(b,b_1),t,r}; f^\textnormal{q}_{(b,b_1),t,r}] \in K:\\& ([\eta_{(b,b_1)}^\textnormal{s}; \eta_{(b,b_1),t,r}^\textnormal{p};\eta_{(b,b_1),t,r}^\textnormal{q}]) \in K^*; \forall b, b_1 \in B
\end{split}\\
\vspace{-10pt}
\label{121}
\begin{split}
&\sum_{b_m \in B_m(b)}\hspace{-10PT}2(X_{(b,b_m)}f^\textnormal{p}_{(b,b_m),t,r} + x_{(b,b_m)}f^\textnormal{q}_{(b,b_m),t,r})\\[-4pt]&+ u_{b,t,r}= \hspace{-5pt}\sum_{b_m \in B_m}\hspace{-5pt}u_{b_{m,t,r}}\hspace{-3pt}: (\beta_{(b,b_m),t,r}); \forall b \in B \} \forall t \in T, r \in R
\end{split}\\
\label{j4}
&e_{y,b}^\textnormal{D} =\hspace{-25PT}\sum_{i \in I(y), t \in T, r \in R}{\hspace{-25PT}R_{y,i}g_{i(b),t,r}}: (\psi_{y,b}^\textnormal{D}); \forall y \in \Theta, b \in B\\
\label{em_d}
&e_{y}^\textnormal{D} = \sum_{b \in B}{e_{y,b}^\textnormal{D}}: (\chi_y^{\textnormal{D}});\hspace{10pt} \forall y \in \Theta\\
&\text{where}\hspace{5pt} \Xi^\textnormal{U}\coloneqq \{d^\textnormal{p}_{b,t,r}, g_{i,t,r}, q_{i,t,r}, e^{\textnormal{D}}_{y}, g_{b_{0},t,r}, q_{b_{0},t,r},\nonumber\\ &\hspace{60pt}f^\textnormal{p}_{(b,b_1),t,r}, f^\textnormal{q}_{(b,b_1),t,r}, u_{b,t,r}\}\nonumber
\end{align}
\end{subequations}
\vspace{-10pt}
\subsection{Demand Modeling}
\label{Consumer}
The nodal demand is modeled using the utility function of aggregated consumers at each node. The utility function represents the response of the aggregated consumers to changes in the electricity tariff, while providing flexibility of modeling the behavior of different consumers using their individual utility function parameters. To accommodate TOU tariffs, we use the Cobb-Douglas utility function where consumer demand at every bus in the system is modeled using two different quantities to represent peak and off-peak electricity prices. The Cobb-Douglas utility function for a representative day $r \in R $ \cite{TOU} is given as follows:
\vspace{-5pt}
\begin{align}
\label{demand}
\begin{split}
&U\Bigg(d^\textnormal{p}_{b,T_1,r} +\sum_{t \in T_1}{D^\textnormal{p}_{b,t,r}} ,  d^\textnormal{p}_{b,T_2,r} +\sum_{t \in T_2}{D^\textnormal{p}_{b,t,r}}\Bigg) \\&=
(d^\textnormal{p}_{b,T_1,r} +\sum_{t \in T_1}{D^\textnormal{p}_{b,t,r}})^{\alpha_{b}} (d^\textnormal{p}_{b,T_2,r}+\sum_{t \in T_2}{D^\textnormal{p}_{b,t,r}})^{1-\alpha_{b}}
\end{split}
\end{align}
where $d^\textnormal{p}_{b,T_1,r}$ is the aggregate flexible demand at bus $b$ during peak demand time $T_1$ for day $r$, $d^\textnormal{p}_{b,T_2,r}$ is the aggregate flexible demand at bus $b$ during off-peak demand time $T_2$, and $\alpha_{b}$ is the demand elasticity parameter at bus ${b}$. Given the peak and off-peak TOU tariffs as $\pi_{b,t,r}^{\textnormal{p}}$ and $\pi_{b,t,r}^{\textnormal{op}}$, the consumer surplus is defined as: 
\vspace{-10pt}

\begin{align}
\label{surplus}
\begin{split}
&\textnormal{CS} =(d^\textnormal{p}_{b,T_1,r} +\sum_{t \in T_1}{D^\textnormal{p}_{b,t,r}})^{\alpha_{b}} (d^\textnormal{p}_{b,T_2,r}+\sum_{t \in T_2}{D^\textnormal{p}_{b,t,r}})^{1-\alpha_{b}}\\& - \pi_{b,t,r}^{\textnormal{p}} (d^\textnormal{p}_{b,T_1,r} +\sum_{t \in T_1}{D^\textnormal{p}_{b,t,r}}) - \pi_{b,t,r}^{\textnormal{op}}(d^\textnormal{p}_{b,T_2,r} +\sum_{t \in T_2}{D^\textnormal{p}_{b,t,r}})
\end{split}
\end{align} 

Similarly, to express peak (off-peak) flexible demand as a function of peak (off-peak) TOU tariffs, we maximize the utility of consumers constrained by a fixed prorated daily budget as below:

\vspace{-10pt}
\begin{subequations}
\label{demand_budget}
\begin{align}
\begin{split}
\max_{\Xi^\textnormal{DM}}&\hspace{2pt} U(d^\textnormal{p}_{b,T_1,r}, d^\textnormal{p}_{b,T_2,r}) = (d^\textnormal{p}_{b,T_1,r})^{\alpha_{b}} (d^\textnormal{p}_{b,T_2,r})^{1-\alpha_{b}}; \\& \hspace{130pt}0 \leq \alpha_{b} \leq 1
\end{split}\\
\label{CD_2}
\begin{split}
\text{s.t.}&\hspace{20pt}
d^\textnormal{p}_{b,T_1,r}\times \pi_{b,t,r}^{\textnormal{p}} + d^\textnormal{p}_{b,T_2,r}\times \pi_{b,t,r}^{\textnormal{op}} \\&\hspace{40pt}=\kappa'\times\mu^\textnormal{inc}_{b}\times \frac{P_{b}}{H_{b}}; \forall b \in B, r \in R
\end{split}\\
\text{where}\hspace{5pt}& \Xi^\textnormal{DM} \coloneqq \{d^\textnormal{p}_{b,T_1,r}, d^\textnormal{p}_{b,T_2,r}, \pi_{b,t,r}^{\textnormal{p}}, \pi_{b,t,r}^{\textnormal{op}} \}\nonumber
\end{align}
\end{subequations}
where the right-hand side in eq.~\eqref{CD_2} defines the fixed budget of consumers  allocated for energy expenses. This constraint is the binding version of eq.~\eqref{r6} in the regulator optimization above,  but the values of $\kappa$ and $\kappa'$ may vary. This difference is to accommodate different priorities of consumers and the regulator for energy expenses.
Next, using the optimality conditions of~\eqref{demand_budget},  we compute:
\begin{subequations}
\begin{align}
\label{demand_peak}
&d^\textnormal{p}_{b,T_2,r} = \frac{(1-\alpha_{b})\kappa'\times\mu^\textnormal{inc}_{b}\times \frac{P_{b}}{H_{b}}}{\pi_{b,t,r}^{\textnormal{op}}};&\hspace{10PT}\forall r \in R\\
\label{demand_offpeak}
&d^\textnormal{p}_{b,T_1,r} = \frac{\alpha_{b}\times
\kappa'\times\mu^\textnormal{inc}_{b}\times \frac{P_{b}}{H_{b}}}{\pi_{b,t,r}^{\textnormal{p}}};&\hspace{10PT}\forall r \in R
\end{align}
\end{subequations}

where $d^\textnormal{p}_{b,T_1,r}$ and $d^\textnormal{p}_{b,T_2,r}$ represent the total demand consumed during peak and off-peak periods at bus $b$ for day $r$. In turn, the power consumed at each interval of these these periods can be calculated as:
\vspace{-10pt}
\begin{subequations}
\begin{align}
\label{demand_peak_t}
&d^\textnormal{p}_{b,t,r} = \overline{d}^\textnormal{p}_{b,t,r}\times \frac{d^\textnormal{p}_{b,T_1,r}}{\overline{d}^\textnormal{p}_{b,T_1,r}}; \forall t \in T_1, r \in R\\
\label{demand_offpeak_t}
&d^\textnormal{p}_{b,t,r} = \overline{d}^\textnormal{p}_{b,t,r}\times \frac{d^\textnormal{p}_{b,T_2,r}}{\overline{d}^\textnormal{p}_{b,T_2,r}}; \forall t \in T_2, r \in R
\end{align}
\end{subequations}
where $\overline{d}^\textnormal{p}_{b,t,r}$, $\overline{d}^\textnormal{p}_{b,T_1,r}$, and $\overline{d}^\textnormal{p}_{b,T_2,r}$ represent the value of inflexible demand at bus $b$ for time interval $t$, $T_1$, and $T_2$ \cite{cobb-douglas}.

\subsection{MOPEC Formulation}
\label{MOPEC_Reform}
To formulate the proposed MOPEC, we first define the SLSF game between the regulator and power utility for designing a justice-cognizant tariff as:
\begin{subequations}
\label{slsf_o}
\begin{align}
\label{slsf_o_1}
\max_{\pi_{b,t}, \Xi^{\textnormal{U}},\Xi^{\textnormal{DM}}} \hspace{10pt}&\textnormal{Eq.}~\eqref{r1}\\
\textnormal{s.t.}\hspace{10pt}
&\textnormal{Eq.}~\eqref{r6} - \eqref{tariff_def}\\
&e_{y,b}^{\textnormal{D}}, d^\textnormal{p}_{b,t,r}, g_{i,t,r} \in \textnormal{arg \{Eq.}~\eqref{distribution}\}\\
& \text{Eq.}~\eqref{demand_peak} - \eqref{demand_offpeak_t}
\end{align}
\end{subequations}

Since eq.~\eqref{slsf_o} is a  bi-level optimization problem with the multi-objective function in eq.~\eqref{slsf_o_1}, it cannot be solved using off-the-shelf solvers.  Thus, we convert this problem into its single-level equivalent using the KKT conditions of the LL (power utility) problem. This single-level formulation contains a multi-objective function and a set of equilibrium constraints, taking the form of a conventional MOPEC, as follows: 

\allowdisplaybreaks
\begin{subequations}
\label{mopec}
\begin{align}
\label{obj}
\max_{\pi_{b,t}, \Xi^{\textnormal{U}},\Xi^{\textnormal{DM}}}\hspace{10pt}& \; \textnormal{Eq.}~\eqref{r1}\\
\label{UL_const}
\text{s.t.}\hspace{10pt}
&\text{UL Constraints:} &\hspace{5pt} \textnormal{Eqs.} & \eqref{r6} -\eqref{tariff_def}
\end{align}
\vspace{-20pt}
\label{KKT_dist}
\begin{align} 
\label{d1}
\begin{split}
&\{(g_{b_{0},t,r}):-\lambda^\textnormal{T}_{t,r}+\lambda_{b_0,t,r}^\textnormal{D} -\underline{\tau}_{(b^\textnormal{T}_c,b_0^\textnormal{D}),t,r}+\overline{\tau}_{(b^\textnormal{T}_c,b_0^\textnormal{D}),t,r}\\&\hspace{100pt}- \zeta_{b_0,t,r}=0; \hspace{3mm}
\end{split}\\
\label{d2}
&(q_{b_{0},t,r}):\lambda_{b_0,t,r}^\textnormal{Dq} =0;\hspace{3mm}\\
\label{d5}
&(q_{i,t,r}):-\lambda_{{b(i),t,r}}^{\textnormal{Dq}}-\underline{\theta}_{i,t,r}+\overline{\theta}_{i,t,r} = 0;\forall i \in I\\
\label{d3}
\begin{split}
&(g_{i,t,r}):-C_{i}+\lambda_{{b(i),t,r}}^{\textnormal{D}}-\underline{\delta}_{i,t,r}+\overline{\delta}_{i,t,r} -\\&\hspace{35pt} \sum_{y \in \theta}{R_{y,i}\psi_{y,b(i)}^\textnormal{D}} = 0;\forall i \in I
\end{split}\\
\label{d6}
\begin{split}
&(f^\textnormal{p}_{(b,b_m),t,r}):-\sum_{b_n \in B_n(b)}{\lambda_{{b_n,t,r}}^{\textnormal{D}}}+\lambda_{{b,t,r}}^{\textnormal{D}}-\\&\hspace{40pt}2\beta_{(b,b_m),t,r}X_{(b,b_m)}-\eta_{(b,b_m),t,r}^\textnormal{p} = 0;\forall b \in B\\
\end{split}\\
\label{d7}
\begin{split}
&(f^\textnormal{q}_{(b,b_m),t,r}):-\sum_{b_n \in B_n(b)}\lambda_{{b_n,t,r}}^{\textnormal{Dq}}+\lambda_{{b,t,r}}^{\textnormal{Dq}}\\&-2\beta_{(b,b_m),t,r}x_{(b,b_m)}-\eta_{(b,b_m),t,r}^\textnormal{q} = 0;\forall b \in B\\
\end{split}\\
\label{d8}
\begin{split}
&(u_{b,t,r}):-\underline{\mu}_{b,t,r}+\overline{\mu}_{b,t,r}+ \sum_{b_m \in B_m(b)}\beta_{(b,b_m),t,r}\\&\hspace{30PT}-\beta_{(b,b_n),t,r}=0;\forall b \in B \};\hspace{5PT}t \in T, r \in R
\end{split}\\
&(e_{y,b}^\textnormal{D}):-\psi_{y,b}^\textnormal{D}+\chi_y =0; \forall y \in \Theta; b \in B\\
&(e_{y}^\textnormal{D}):\chi_y^{\textnormal{D}} =0; \forall y \in \Theta/CO_2\\
&(e_{CO_2}^\textnormal{D}):-\gamma- \chi_{CO_2}^{\textnormal{D}} =0; \\
\label{d9}
&0 \leq g_{i,t,r}-G_{{i}}^\textnormal{{min}} \;\bot\; \underline{\delta}_{i,t,r} \geq 0;\forall i \in I, t \in T, r \in R\\
\label{d10}
&0 \leq G_{{i}}^\textnormal{{max}}-g_{i,t,r} \;\bot\; \overline{\delta}_{i,t,r} \geq 0;\forall i \in I, t \in T, r \in R\\
\label{d11}
&0 \leq q_{i,t,r}-Q_{{i}}^\textnormal{{min}} \;\bot\; \underline{\theta}_{i,t,r} \geq 0;\forall i \in I, t \in T, r \in R\\
\label{d12}
&0 \leq Q_{{i}}^\textnormal{{min}}-q_{i,t,r} \;\bot\; \overline{\theta}_{i,t,r} \geq 0;\forall i \in I, t \in T, r \in R\\
\label{d13}
&0 \leq u_{b,t,r}-U_{{b}}^\textnormal{{min}} \;\bot\; \underline{\mu}_{b,t,r} \geq 0;\forall b \in B, t \in T, r \in R\\
\label{d14}
&0 \leq U_{{b}}^\textnormal{{max}}-u_{b,t,r} \;\bot\; \overline{\mu}_{b,t,r} \geq 0;\forall b \in B, t \in T, r \in R\\
\label{d14.1}
&0 \leq g_{b_{0},t,r} \;\bot\;\underline{\tau}_{(b^\textnormal{T}_c,b_0^\textnormal{D}),t,r} \geq 0; \forall t \in T, r \in R\\ 
\label{d14.2}
&0 \leq F_{(b^\textnormal{T}_c,b_0^\textnormal{D})}^{\textnormal{max}} -g_{b_{0},t,r} \;\bot\; \overline{\tau}_{(b^\textnormal{T}_c,b_0^\textnormal{D}),t,r} \geq 0; \forall t \in T, r \in R\\
\begin{split}
&[S_{(b,b_1)}; f^\textnormal{p}_{(b,b_1),t,r}; f^\textnormal{q}_{(b,b_1),t,r}] \;\bot\;[\eta_{(b,b_1)}^\textnormal{s}; \eta_{(b,b_1),t,r}^\textnormal{p};\\&\hspace{50pt}\eta_{(b,b_1),t,r}^\textnormal{q}]; \forall b, b_1 \in B, t \in T, r \in R
\end{split}\\
\label{d15}
&\text{Equality Constraints: Eqs.}~\eqref{7} - \eqref{91}, \eqref{121} - \eqref{em_d}, \eqref{demand_peak} - \eqref{demand_offpeak_t}
\end{align}
\end{subequations}

where eq.~\eqref{obj} is the multi-objective function of the UL (regulator) problem, and eq.~\eqref{UL_const} is the set of UL constraints. Eqs.~\eqref{d1} - \eqref{d15} are the KKT conditions of the LL (power utility) problem.


\section{Solution Technique}
In this section, we detail the solution algorithm employed to solve the MOPEC defined in eq.~\eqref{mopec}. First, we introduce  a generic equivalent of the MOPEC in eq.~\eqref{mopec}, which simplifies algorithmic discussions below:
\begin{align}
    \min_{\bold{x,y}}\;\bold{F(x,y)} = &[F_1(\bold{x,y}),F_2(\bold{x,y}),\dots, F_k(\bold{x,y})]^\textnormal{T}\\
    \text{s.t.}\; &(\bold{x,y}) \in \Psi\\
    & \bold{y} \in \bold{S(x)}
\end{align}
where $\bold{F(x,y)}$ is the multi-objective function of this MOPEC (denoted as $\textnormal{O}$ in eq.~\eqref{r1}) and $F_i(\bold{x,y})$ are the components of $\bold{F(x,y)}$.  Vectors $\bold{x}$ and $\bold{y}$ are the sets of UL and LL decision variables, while $\Psi$ is the joint feasible region, i.e., between the problem of the regulator and problem of the power utility,  such that the optimal solution exists in the intersecting space between the individual feasible regions of these two problems. Similarly, $\bold{S(x)}$  is the solution set of the LL problem, which  is a non-linear complementarity problem (NCP) due to the  non-linear and complementarity LL constraints \eqref{d1} - \eqref{d15}.  Hence, solving this NCP for a vector $\bold{g}$ implies finding a vector $\bold{z}$ such that:
\vspace{-5pt}
\begin{subequations}
\label{comp_prob}
\begin{align}
    &\bold{z}\geq 0 \; ; \; \bold{g(z)}\geq0;\\
    &\bold{z^Tg(z)}=0
\end{align}
\end{subequations}
where $\bold{z} \subseteq \bold{y}$, and eq.~\eqref{comp_prob} $\subseteq \bold{S(x)}$. Hence, a generic NCP-based MOPEC can be written as:
\begin{subequations}
\label{MOPEC_NCP}
\begin{align}
    \min_{\bold{x,y}}\;\bold{F(x,y)} = &[F_1(\bold{x,y}),F_2(\bold{x,y}),\dots, F_k(\bold{x,y})]^\textnormal{T}\\
    \text{s.t.}\; &(\bold{x,y}) \in \Psi\\
    &\bold{y}\geq 0 \\
    &\bold{g(x,y)}\geq0\\
    \label{mopec_comp}
    &\bold{y^Tg(x,y)}=0.
\end{align}
\end{subequations}

\begin{figure}[!t]
\centering
\includegraphics[height=2in, width=\columnwidth]{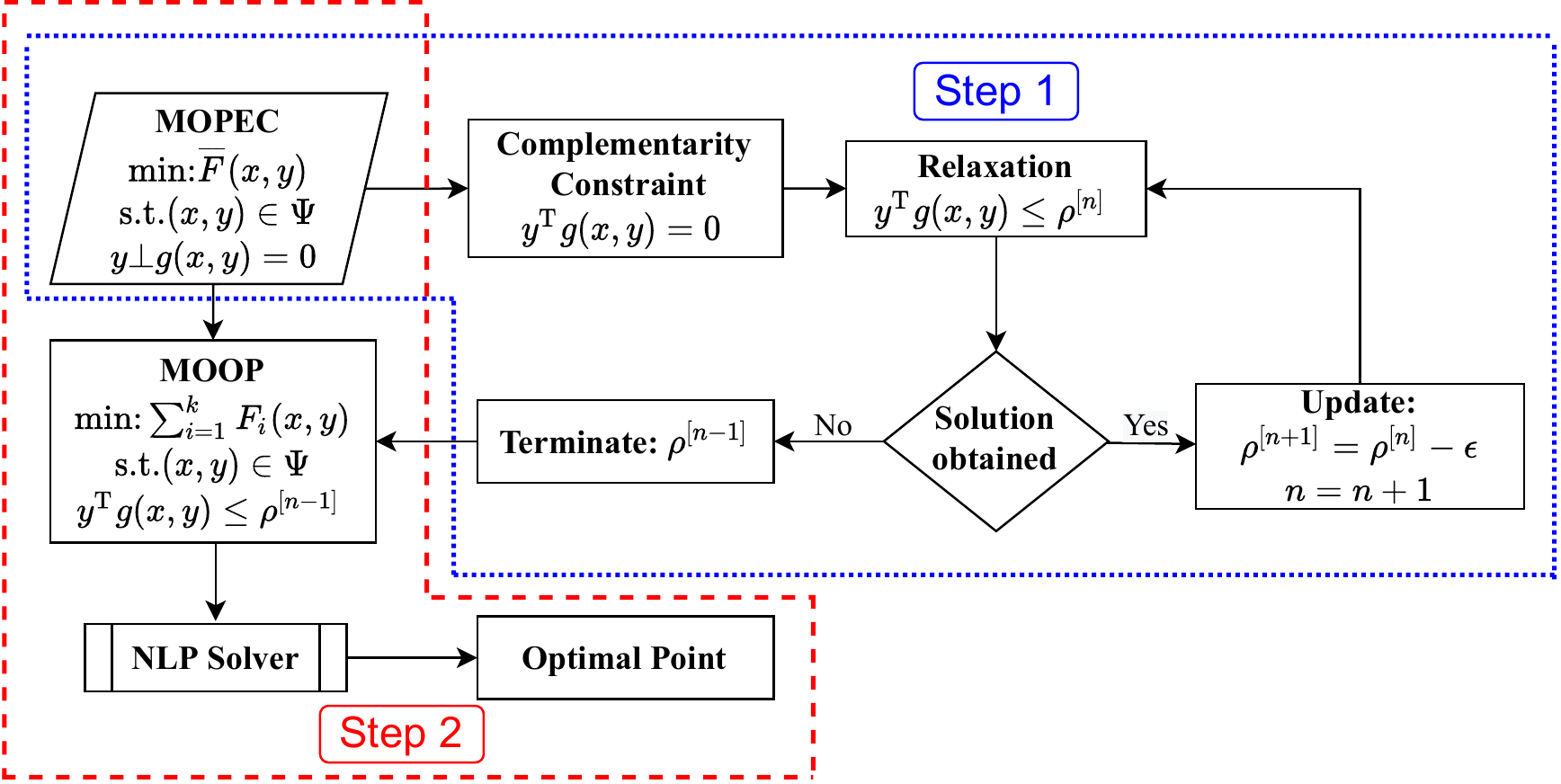}
\caption{\small{Flowchart for the proposed integrated MOPEC solution technique. Step 1 (dotted blue box) is the Scholtes's relaxation for complementarity constraints and Step 2 (dashed red box) is the objective sum method for MOOP.}}
\label{flowchart}
\vspace{-20pt}
\end{figure}
\vspace{-5pt}
The complementarity constraints in eq.~\eqref{MOPEC_NCP} form a disjoint, non-convex feasible region \cite{MOPEC_stat}, which precludes  obtaining a global optimal solution for this MOPEC.  To overcome this complexity, we propose the integration of solution methods for multi-objective optimization problems and MPECs. The resulting integrated solution methodology for MOPECs is shown in Fig.~\ref{flowchart}, where Step 1 is the treatment of complementarity constraints using Scholtes's relaxation to convert the MOPEC into a multi-objective optimization problem (MOOP). This MOOP is then reformulated using the objective sum method in Step 2, which is finally solved with an off-the-shelf NLP solver.  This methodology generates several Pareto optimal points (frontier), while applying the MPEC solution technique only once, and is achieved in two steps described below.

\subsubsection{MPEC Solution Technique for Complementarity Constraints in MOPEC}
For the complementarity problem defined in eq.~\eqref{comp_prob}, we employ the global relaxation scheme by Scholtes \cite{Scholtes}. This relaxation makes it possible to reformulate complementarity constraints, defined in eq.~\eqref{mopec_comp}, as $\bold{\bold{y}^\textnormal{T} \bold{g(x,y)}} \leq \rho$, where $\rho$ is a small, user-defined parameter. The resulting NLP formulation is iteratively solved such that in each iteration, the value of $\rho$ is decreased from its previous value till $\rho \rightarrow 0$. The associated solution of the relaxed NLP, where $\rho \approx 0$, is the solution of the complementarity problem. With the application of Scholtes's relaxation technique, the resulting formulation does not contain strict complementarity constraints, and resembles a generic non-linear MOOP.
\subsubsection{Solution technique for MOOP}
For the non-linear MOOP attained from the application of Scholtes's relaxation technique, we employ the objective sum method as follows. For a generic MOOP defined as:
\begin{subequations}
\begin{align}
    \min_{x}\;\bold{F(x)} = &[F_1(\bold{x}),F_2(\bold{x}),\dots, F_k(\bold{x})]^\textnormal{T}\\
    \text{s.t.}\;&g_j(x)\geq0 \; \forall j \\
    &h_i(x)=0 \; \forall i
\end{align}
\end{subequations}
the objective sum formulation can be written as:
\vspace{-5pt}
\begin{subequations}
\label{sum_obj}
\begin{align}
\label{sum_obj_1}
    \min_{x}\;\bold{F(x)} = &\sum_{n=1}^{k}{\omega_n F_n(\bold{x})}\\
    \text{s.t.}\;&g_j(x)\geq0 \; \forall j \\
    &h_i(x)=0 \; \forall i
\end{align}
\end{subequations}

where $\omega_n$ is the weight associated with the n\textsuperscript{th} component of the multi-objective function $\bold{F(x)}$. These weights are chosen by the policy maker (regulator in this case), based on the articulation of their preferences for different objectives in the multi-objective function, leading to the concept of a generic preference function. The choice of weight vector $\omega$ affects the obtained optimal solution, since ideally, a different weight vector should result in a different optimal point on the Pareto frontier. However, in reality this conjecture does not always hold, resulting in a non-uniform distribution of optimal points on the Pareto frontier \cite{MOOP_book}.

\section{Case Study}
\begin{figure}[!t]
\centering
\includegraphics[width=3.5in, height=2.8in]{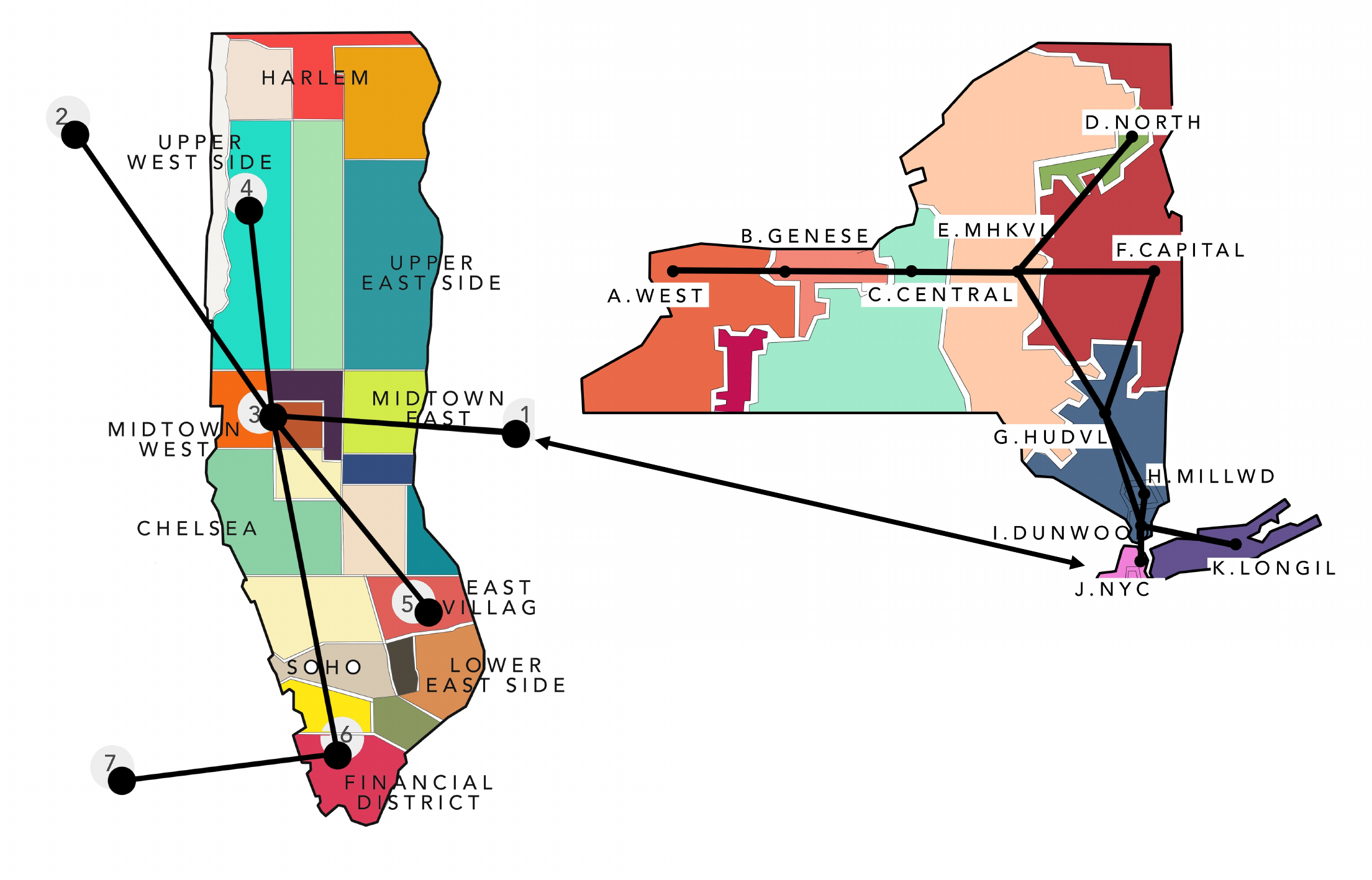}
\vspace{-30pt}
\caption{\small{A diagram of the 11-zone NYISO transmission system connected to the 7-bus Manhattan distribution system. The T\&D interconnection is shown between NYC and bus \# 1.
}}
\label{Manhattan}
\vspace{-10pt}
\end{figure}

The case study uses  the 11-zone NYISO transmission system \cite{SLMF} for the wholesale market and the 7-bus Manhattan distribution system \cite{samrat} for the power utility. Fig.~\ref{Manhattan} displays the Manhattan and NYISO systems. The interconnecting line between the T\&D networks is modeled between bus \# 1 in Manhattan and bus \# 10 (NYC) in the NYISO system. Simulations are carried out using the data available in \cite{goldbook}. The socio-demographic data used for the Manhattan system is shown in Table~\ref{socio_data} \cite{income_data}. We use the minimum of the median household income of zip codes at each bus of the Manhattan system to generate the minimum average household income. This metric of household income ensures that no household is overburdened with the designed tariff. Conversely, using median household income of all zip codes would overburden households with less than the median income.

\begin{center}
\begin{table}[!t]
\caption{Demographic data for the  Manhattan system, \cite{income_data}.}
\centering
\begin{tabular}{ >{\centering}p{0.5cm} | >{\centering}p{2.5cm} | >{\centering}p{1.1cm} | >{\centering\arraybackslash}p{2cm}}
\hline
\hline
Bus No.  & Min. Avg. Household Income (\$) & Population & Maximum Load (MW)  \\
\hline 
3 & 38,304 & 206,707 & 742 \\
4& 14,896 & 479,911 & 855 \\
5& 24,022 & 596,438 & 314 \\
6 & 31,316 & 29,266 & 188\\
\hline
\end{tabular}
\label{socio_data}
\vspace{-15pt}
\end{table}
\end{center}
\vspace{-20pt}
Using this data and the formulation  in Section~\ref{MOPEC_Reform}, we define Case 1 as the justice-cognizant tariff design framework with flat system-wide tariff, whereas Case 2 solves the same problem for the TOU tariff. The range of the peak and off-peak hours used in this case study is set based on peak and off-peak load hours determined by Con Edison \cite{tariff}. Case 3 extends Case 2 by introducing a distinct TOU tariff for each bus in the Manhattan system. Case 4 builds on  Case 3 by adding hourly time granularity to the Case 3 tariff. Case 4 is different from real-time pricing discussed in literature \cite{Ansarin}, since in addition to the hourly variation in tariff, we also allow a spatial variation to cater for different socio-economic and demographic features in the system. We vary the percentage of EB between 6\% and 12\% to accommodate the range defined in \cite{ACEE} for mean to severe household EBs, and observe its effects on tariff, recoverable capital cost of utility, and the value of multi-objective function ($\textnormal{O}$) in eq.~\eqref{r1}. Although in practice the regulator aims to reduce the EB in the system, we evaluate all the cases for different EB values to compare their effect on economic, environmental and public health impacts generated under different tariff structures. An 11\% rate of return on capital investment, available in \cite{ConEd_Report}, is chosen for the utility \cite{AEE_portal}. The capital cost is prorated on a daily basis, while 20 years of cost recovery with a 5\% discount rate is used. Pollutant emission factor values ($R_{y,i}$) are obtained from \cite{Felder, EPA}, while carbon tax ($\gamma$) and Social Cost of Carbon ($\gamma^\textnormal{sc}$) are available in \cite{SCC}. Unless stated otherwise,  $\omega = [1, 1, 1]$ for eq.~\eqref{sum_obj_1}.  The case study is implemented in JuMP v0.20 using Ipopt v3.12.10 on an Intel Core i5 2.4 GHz processor with 4 cores and 8GB of memory. 
\vspace{-25pt}
\begin{center}
\begin{table}[!t]
\caption{Optimal flat system-wide tariff for different EB values (Case 1)}
\centering
\begin{tabular}{ >{\centering}p{1.5cm} | >{\centering}p{1cm} |>{\centering}p{2.5cm} |>{\centering\arraybackslash}p{1.5cm}}
\hline
\hline
Energy Burden  & Tariff [\$/MWh] & Multi-objective function (\textnormal{O})[\$] & Profit of the Utility [\$]  \\
\hline 
6\%--8\% & \multicolumn{3}{c}{Infeasible}\\
\hline
9\% & 39.26 & 1.283$\times 10^9$ & 3.46$\times 10^5$\\
10\%--12\% & 42.68 &  1.254$\times 10^9$ & 3.63$\times 10^5$\\
\hline
\end{tabular}
\label{Case 1}
\vspace{-20pt}
\end{table}
\end{center}

\subsection{Case 1: Flat System-wide Tariff}
Table~\ref{Case 1} summarizes the results for Case 1, where a flat system-wide tariff is considered. For the EB of 6\%--8\%, which is a parameter set by the regulator before solving the proposed optimization, the utility cannot fully recover its capital cost, which renders infeasibility due to a low revenue collected from the tariff. For the EB of 9\%--12\%, the utility  recovers its full capital cost, making this EB range feasible for the regulator to optimize its policy decisions. As expected, value of the tariff increases with relaxing the EB constraint, however, remains constant from 10\% to 12\%. This is because the EB constraint (eq.~\eqref{r6}) is binding for all values of EB $\leq 10\%$, determining the value of tariff. However, for EB $> 10\%$, eq.~\eqref{r6} is non-binding, therefore, the value of tariff does not increase owing to eq.~\eqref{r8}. Thus, the EB of 10\% provides the minimum possible value of objective function $\textnormal{O}$. 
Notably, the decrease in the value of $\textnormal{O}$ as the EB increases from 9\% to 10\% is counter-intuitive. However, while increasing the EB increases the tariff,  it only slightly reduces the utility of consumers, while greatly increasing the operational cost of the power utility. The operational cost increases since in this case oil and gas generators in Zone J of NYISO system  with significantly high locational marginal external costs but slightly low operational cost cannot be dispatched. Therefore, similar generators  outside Zone J with lower marginal external costs (though still greater than a typical, non-marginal natural gas unit) but higher operational costs are dispatched.  As a result, relaxing the EB constraint to 10\% increases $f^\textnormal{EW}$ , whereas $f^\textnormal{H}$ and $f^\textnormal{EN}$ decrease, resulting in an overall lower value of $\textnormal{O}$. The constant values of $\textnormal{O}$ for the EB of  10\%--12\% are observed due to achieving the maximum tariff value and minimum possible value of $\textnormal{O}$. Similarly, profit of the utility increases as the EB increases, however, remains constant for the EB of 10\%--12\% due to a constant tariff in this range.

\vspace{-25pt}
\begin{center}
\begin{table}[!t]
\caption{Optimal time-of-use tariff for different EB values (Case 2)}
\centering
\renewcommand{\arraystretch}{1}
\begin{tabular}{ >{\centering}p{1cm} | >{\centering}p{1.3cm} | >{\centering}p{1.8cm} | >{\centering}p{2.0cm} | >{\centering\arraybackslash}p{1.4cm}}
\hline
\hline
Energy Burden  & Peak Tariff [\$/MWh] & Off-Peak Tariff [\$/MWh] & Multi-objective function (\textnormal{O})[\$] & Profit of the Utility [\$] \\
\hline 
6\%--7\% & \multicolumn{4}{c}{Infeasible}\\
\hline
8\% & 52.98 & 16.8 & 1.290$\times 10^9$ & 2.40$\times 10^5$\\
9\% & 61.72 & 16.8  & 1.254$\times 10^9$ & 4.76$\times 10^5$ \\
10\% & 59.41 & 23.21 & 1.254$\times 10^9$ & 5.15$\times 10^5$ \\
11\% & 57.46 & 25.53 & 1.254$\times 10^9$ & 5.15$\times 10^5$\\
12\% & 55.49 & 27.86 & 1.254$\times 10^9$ & 5.15$\times 10^5$\\
\hline
\end{tabular}
\label{Case 2}
\vspace{-10pt}
\end{table}
\end{center}

\subsection{Case 2: TOU Tariff}
Table~\ref{Case 2} shows the results for Case 2 where the TOU tariff is considered. The power utility cannot recover its capital cost for the  EB of 6\%--7\%, whereas for the EB of 8\%--12\%, the power utility fully recovers its capital cost. As the EB increases from 8\% to 10\%, the off-peak tariff remains constant and the peak tariff increases as expected. This is coupled with a decrease in the value of objective function $\textnormal{O}$ due to a significant increase in the operational cost of the utility, as explained in Case 1. However, for the EB of 9\%--12\%, the values of the peak tariff decrease accompanied by a simultaneous increase in the values of the off-peak tariff, and a constant value of $\textnormal{O}$. This trend can be explained by eq.~\eqref{r6}, which is binding in this case only for the EB $\leq 9\%$. Since, for EB $> 9\%$, eq.~\eqref{r6} is non-binding, the optimization problem chooses such values of peak and off-peak tariffs that keep the average system tariff constant and do not over-burden consumers with high peak tariffs. This flexibility is not available when eq.~\eqref{r6} is binding since the minimum value of off-peak tariff, defined in eq.~\eqref{r9}, is chosen as the off-peak tariff. Profit of the utility increases as the EB increases, however, remains constant for EB $\geq 10\%$. The constant tariff in this range is because the average system tariff, as defined in eq.~\eqref{r11}, remains constant for EB $\geq 10\%$. 

Relative to Case 1,  the feasible EB range  increases and the minimum achievable value of $\textnormal{O}$ shifts to the EB of 9\% in Case 2. Hence, similar economic, environmental, and health benefits can be achieved for a lower EB with a TOU tariff as compared to the flat system-wide tariff.

\subsection{Case 3: Locational TOU Tariff}
\begin{figure}[!t]
\centering
\includegraphics[scale=0.18]{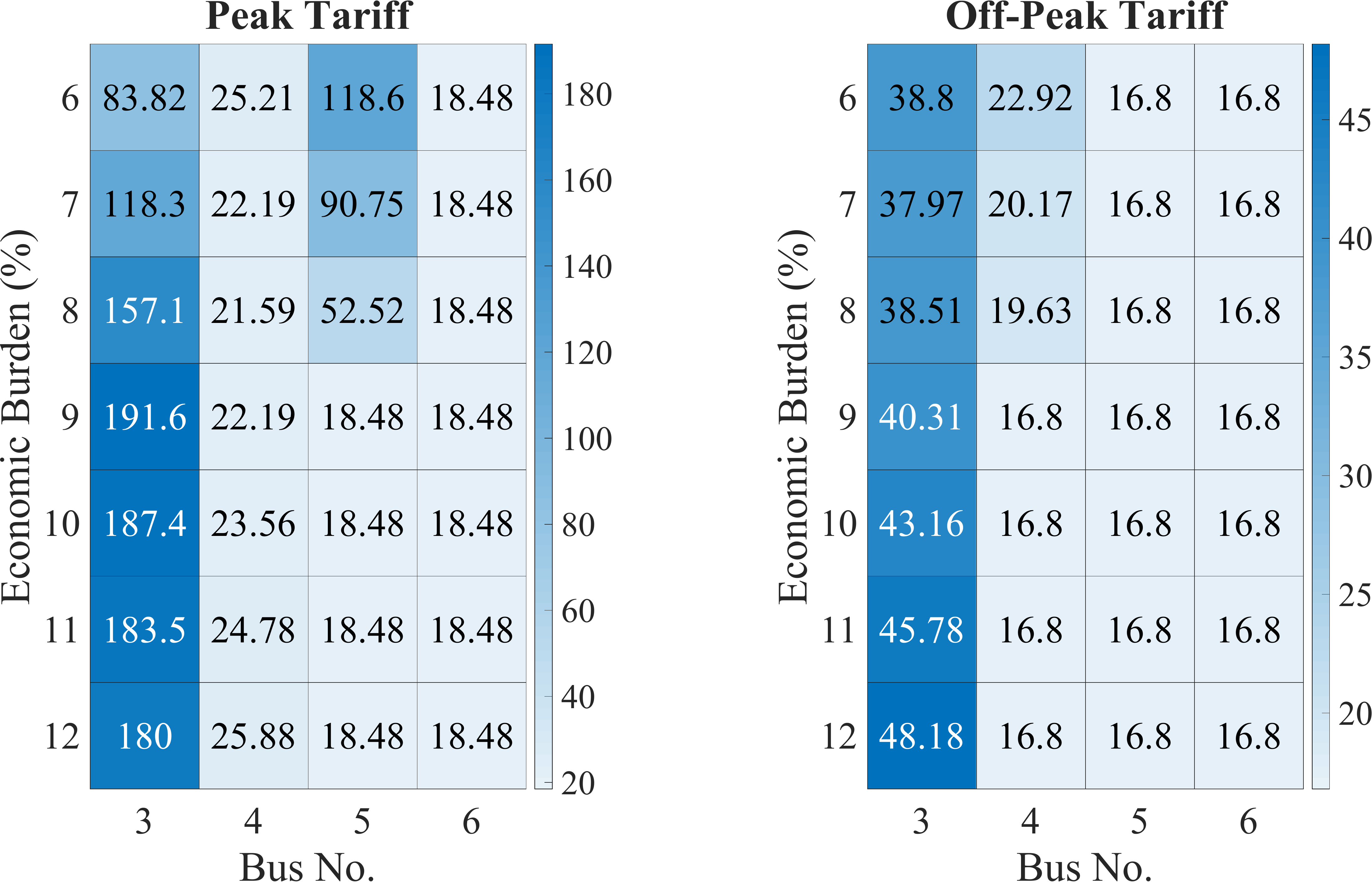}
\caption{\small{Locational TOU Tariff (Case 3): Variations in peak and off-peak tariffs at each bus for different EB levels.}}
\label{Case 3}
\vspace{-10pt}
\end{figure}

Fig.~\ref{Case 3} shows the results for Case 3 where a locational TOU tariff is considered, i.e. a unique TOU tariff is defined for each bus. From Fig.~\ref{Case 3}, the highest peak tariffs are observed for bus \# 3, followed by bus \# 5 for the EB of 6\%--8\%, and bus \# 4 for the EB of 9\%--12\%. The minimum peak tariffs are observed for bus \# 6. Similarly, bus \# 3 has the highest off-peak tariffs, followed by bus \# 4 for the EB of  6\%--8\%, whereas the off-peak tariffs for buses \# 4, 5, and 6 for the EB of 9\%--12\% remain constant. Correlating these results with the demographic features in Table~\ref{socio_data}, we see that bus \# 3 has the highest minimum average household income and the highest tariffs, followed by bus \# 5 which has the third highest household income but the highest population. Bus \# 6 has the lowest peak tariff although its household income is higher than bus \# 4 and bus \# 5, however, it has a significantly low population and total load. Hence, tariffs are not only a function of household income, but the population and total load also impact tariff values.  The highest average tariffs, as defined in eq.~\eqref{r11},  are observed for bus \# 3, followed by bus \# 5 for the EB of 6\%--8\%, and bus \# 4 for the EB of 9\%--12\%. This trend is consistent (with the exception of bus \# 6) with the trend in minimum average household income on these buses. However, as the EB increases from 9\% to 12\%, bus \# 4 is burdened more, even with a relatively low household income. This is because the increase in the EB on bus \# 4 allows the optimization framework to choose higher tariffs at this bus (owing to a lower demand flexibility), while simultaneously reducing the tariffs at bus \# 5 due to its high load flexibility. 

The value of objective function $\textnormal{O}$ for all EB values in Case 3 remains constant and  the capital cost of the power utility is also fully recovered. Comparing this with Cases 1 and 2 shows that maximum benefits for the power utility and consumers can be yielded with an EB of 6\%. Hence, the economic efficiency of this tariff is higher and the net social benefits are distributed more equitably as compared to Cases 1 and 2.

\subsection{Case 4: Locational Hourly Tariff}
\begin{figure}[!t]
\centering
\includegraphics[scale=0.22]{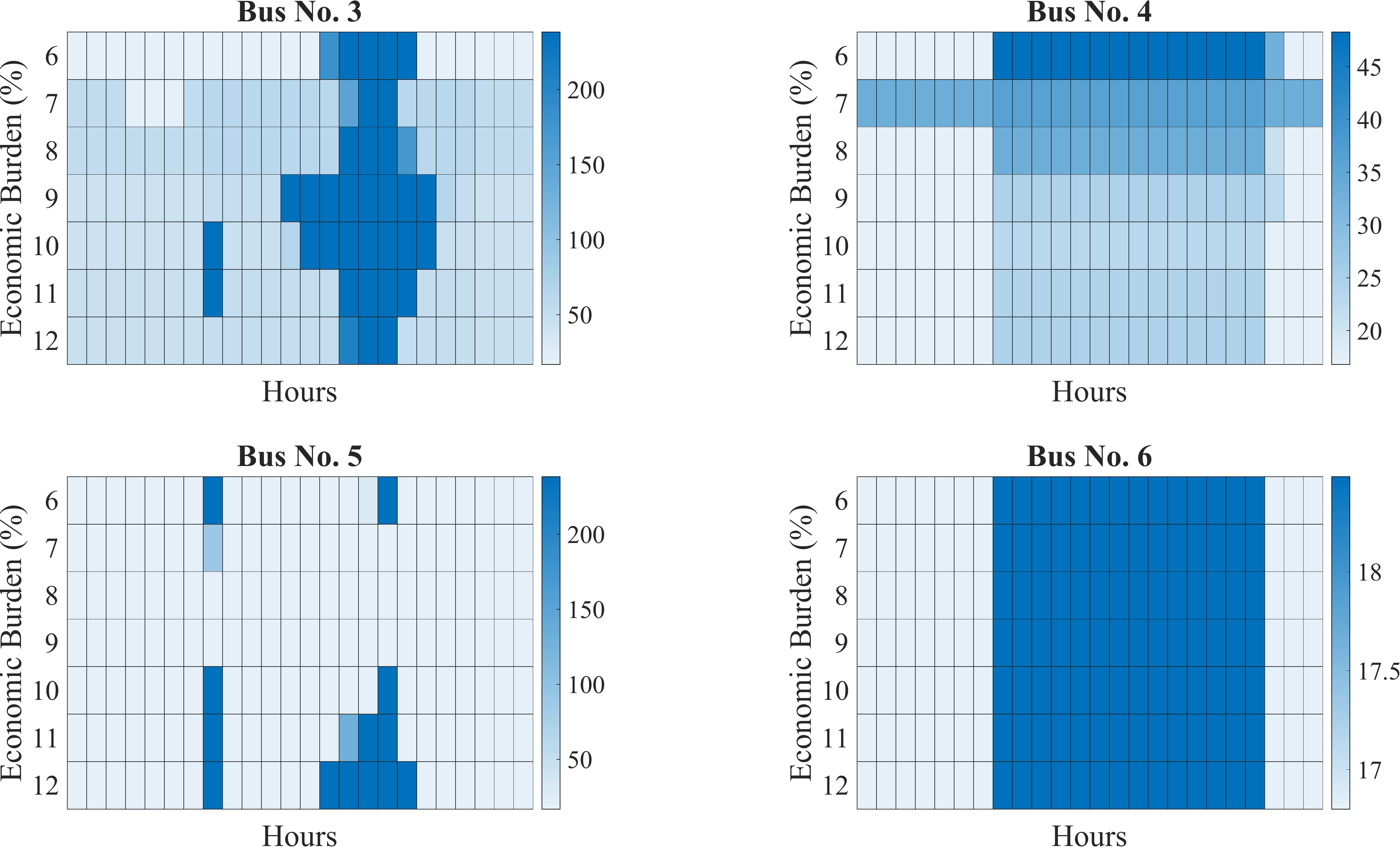}
\caption{\small{Locational Hourly Tariff (Case 4): Hourly tariff variations at each bus for different EB levels. }}
\label{Case 4}
\vspace{-10pt}
\end{figure}

Fig.~\ref{Case 4} shows the results for Case 4 where a locational hourly tariff is considered. The tariffs at each bus follow the trends similar to  Case 3, i.e. the highest tariff is at bus \# 3 followed by bus \# 5 and bus \# 4. However, a finer time resolution allows for higher  tariffs at bus \# 3 and bus \# 4 during the maximum load hours (3pm - 5pm), where significantly lower tariffs are observed otherwise. Moreover, during the maximum load hours for the EB of 6\%--9\%, tariffs at bus \# 5 are not as high as for the EB of 10\%--12\%. Comparing these values to bus \# 3 and bus \# 4, we see that a small tariff decrease  at bus \# 5 is accompanied by a slight tariff increase at bus \# 4, and a considerable tariff increase at bus \# 3. However, the average system tariff remains within bounds, defined in eq.~\eqref{r11}. Similarly, for all EB values and all time periods, the resulting tariff at bus \# 4 is less than the tariff at bus \# 5, which is consistent with the household income at these two buses. Contrasting this with Case 3, the EB range does not impact the trend of tariff in Case 4. Hence, for all EB values, a high temporal resolution allows for choosing tariffs consistent with household incomes  at all times. 

Similar to Case 3, the power utility in this case can fully recover its capital cost at the minimum value of objective function  $\textnormal{O}$ for all EB values. However, unlike Case 3, the individual tariff values  at bus \# 4 and bus \# 5 are  consistent with the household income. Hence, this tariff is economically as efficient as Case 3, however, more equitable in terms of allocation of costs and benefits in the system.

\subsection{Articulation of Preferences in the Multi-Objective Function}

\begin{center}
\begin{table}[!t]
\caption{Effects of articulating  preferences of the regulator}
\centering
\renewcommand{\arraystretch}{1}
\begin{tabular}{>{\centering}p{0.4cm}|
>{\centering}p{1.2cm} | >{\centering}p{1.3cm} | >{\centering}p{1.8cm} | >{\centering}p{0.7cm} | >{\centering\arraybackslash}p{0.7cm}}
\hline
\hline
\multirow{2}{0.4cm}{Case No.} &\multirow{2}{1.2cm}{Energy Burden}  & \multirow{2}{1.3cm}{Peak Tariff [\$/MWh]} & \multirow{2}{1.8cm}{Off-Peak Tariff [\$/MWh]} & \multicolumn{2}{c}{\parbox[t]{2.2cm}{Multi-Objective Func. (\textnormal{O})[$\times 10^9$\$]}}  \\\cline{5-6}
&&&&$\omega=[1,2,2]$ & $\omega=[1,5,5]$\\
\hline 
\multirow{2}{*}{1}&9\% & \multicolumn{2}{c|}{41.00}& 2.564 & 6.41\\

&10\%--12\% & \multicolumn{2}{c|}{42.68}& 2.508 & 6.27\\
\hline 
 
\multirow{2}{*}{2} & 7\% & 64.18 & 16.8 & 2.512 & 6.28\\\cline{3-4}

& 8\%--12\% & \multicolumn{2}{c|}{Same as in Table~\ref{Case 2}} & 2.508& 6.27\\
\hline
\end{tabular}
\label{MO}
\vspace{-15pt}
\end{table}
\end{center}
\vspace{-15pt}

To articulate the justice preferences of the regulator, this section considers $\omega = [1,2,2]$ and $\omega = [1,5,5]$, that is a greater weight is placed on terms modeling  environmental ($f^{\text{EN}}$)  and public health ($f^{\text{H}}$) considerations relative to economic welfare ($f^{\text{EW}}$). In both instances, the tariff trends for Cases 1--4 remain the same. Relative to $\omega = [1,1,1]$, tariff in Case 1 remains the same for the EB of 10\%--12\%, however, for the EB of 9\%, tariff increases from 39.26 \$/MWh to 41.0 \$/MWh. This increase in tariff and reduction in the value of objective function \textnormal{O}, as shown in Table~\ref{MO}, is because the objective function value can still be reduced by increasing the tariff, which is caused by the dispatch of more expensive generators with less public health impacts due to their location. Similarly, for Case 2, relative to  $\omega = [1,1,1]$, the feasible EB range increases to 7\%--12\%. Again, a lower preference for economic welfare allows the dispatch of more expensive generators, increasing the tariff and hence the revenue of the power utility, which results in an increased feasible EB range. The minimum value of  objective function $\textnormal{O}$ is achieved for the EB of 8\%--12\%, and the tariff remains the same as for $\omega = [1,1,1]$. For Cases 3 and 4, the minimum objective function value is achieved for the EB of 6\%--12\%, and the tariff remains the same as for $\omega = [1,1,1]$ throughout this range. Thus, similar to the analysis with $\omega=[1,1,1]$, hourly locational tariff is the most equitable tariff structure to distribute costs and benefits in the system irrespective of the articulation of preferences by the regulator. The tariff values on each bus in this case are consistent with the mean household income, and minimum possible environmental and public health damages are incurred for the EB of 6\%.

\section{Conclusion}
This paper proposes a justice-cognizant tariff design framework that incorporates the previously ignored socio-demographic dimensions in this problem. These dimensions, motivated by the energy justice framework, include economic welfare, public health, and environmental impacts of the designed tariff, along with the mean household income of consumers.  The proposed framework is applied on four different tariff structures, and the results demonstrate that the current and widely-used flat system-wide tariff and TOU tariff not only over-burden the consumers from an economic perspective, but also compromise the environment and public health in an attempt to lower the EB. For example, the lowest feasible EB values for Cases 1 (Flat System-wide Tariff) and 2 (TOU Tariff) are 9\% and 8\% respectively, but they result in higher environmental and public health damages. On the other hand, for Cases 3 (Locational TOU Tariff) and 4 (Locational Hourly Tariff), the EB of 6\% results in the minimum achievable environmental and public health degradation. Hence, a justice-cognizant, and temporally and spatially granular optimal tariff simultaneously improves the economic efficiency and equity in the system.

\bibliographystyle{IEEEtran}
\bibliography{references}{}

\end{document}